\documentstyle[prd,aps,preprint,tighten,epsfig]{revtex}

\begin{document}

\draft

\title{A full parametrization of the $6\times 6$ flavor mixing
matrix \\ in the presence of three light or heavy sterile neutrinos}
\author{{\bf Zhi-zhong Xing}
\thanks{E-mail: xingzz@ihep.ac.cn}}
\address{\sl Institute of High Energy Physics, Chinese Academy of
Sciences, Beijing 100049, China}
\maketitle

\begin{abstract}
In addition to three active neutrinos $\nu^{}_e$, $\nu^{}_\mu$ and $\nu^{}_\tau$,
one or more light sterile neutrinos have been conjectured to account for
the LSND, MiniBooNE and reactor antineutrino anomalies (at the sub-eV mass
scale) or for warm dark matter in the Universe (at the keV mass
scale). Heavy Majorana neutrinos at or above the TeV scale have also
been assumed in some seesaw models. Such hypothetical particles can
weakly mix with active neutrinos, and thus their existence can
be detected at low energies. In the (3+3) scenario with three sterile
neutrinos we present a full parametrization of the $6\times 6$
flavor mixing matrix in terms of fifteen rotation angles and fifteen
phase angles. We show that this standard parametrization allows
us to clearly describe the salient features of some problems in
neutrino phenomenology, such as (a) possible
contributions of light sterile neutrinos to the tritium beta decay
and neutrinoless double-beta decay; (b) leptonic CP violation
and deformed unitarity triangles of the $3\times 3$ flavor
mixing matrix of three active neutrinos; (c) a reconstruction of the
$6\times 6$ neutrino mass matrix in the type-(I+II) seesaw mechanism;
and (d) flavored and unflavored leptogenesis scenarios in the type-I
seesaw mechanism with three heavy Majorana neutrinos.
\end{abstract}


\newpage

\section{Introduction}

One of the fundamental questions in neutrino physics and cosmology
is whether there exist extra species of neutrinos which do not
directly participate in the standard weak interactions.
Such sterile neutrinos are certainly hypothetical,
but their possible existence is either theoretically motivated or
experimentally implied. For example,
\begin{itemize}
\item     heavy Majorana neutrinos at or above the TeV scale
are expected in many seesaw models \cite{SS},
which can not only interpret the small masses of three active neutrinos
but also account for the cosmological matter-antimatter asymmetry via
the leptogenesis mechanism \cite{FY};

\item     the LSND antineutrino anomaly \cite{LSND}, the MiniBooNE
antineutrino anomaly \cite{M} and the reactor antineutrino
anomaly \cite{R} can all be explained as the active-sterile
antineutrino oscillations in the assumption of two species of sterile
antineutrinos whose masses are close to 1 eV \cite{Schwetz};

\item     an analysis of the existing data on the cosmic microwave
background (CMB), galaxy clustering and supernovae Ia favors some
extra radiation content in the Universe and one or two species of
sterile neutrinos at the sub-eV mass scale
\cite{Raffelt}
\footnote{If the bound obtained from the Big Bang nucleosynthesis is
taken into account, however, only one species of light sterile neutrinos
and antineutrinos is allowed \cite{Mangano}.};

\item     sufficiently long-lived sterile neutrinos in the keV mass
range can serve for a good candidate for warm dark matter, whose
presence may allow us to solve or soften several problems that we
have recently encountered in the dark matter simulations \cite{Bode} (e.g.,
to damp the inhomogeneities on small scales by reducing the number
of dwarf galaxies or to smooth the cusps in the dark matter halos)
\footnote{There are some interesting models
which can accommodate sterile neutrinos at either keV
\cite{keV} or sub-eV \cite{eV} mass scales. A model-independent
argument is also supporting the conjecture of warm dark matter particles
hiding out in the ``flavor desert" of the fermion mass spectrum
\cite{Desert}.}.
\end{itemize}
No matter how small or how large the mass scale of sterile neutrinos
is, they are undetectable unless they mix with three active neutrinos
to some extent. The strength of active-sterile neutrino
mixing can be described in terms of some rotation angles and
phase angles, just like the parametrization of the $3\times 3$ quark
flavor mixing in the standard model \cite{PDG}.

The main purpose of this paper is to present a full parametrization
of the $6\times 6$ flavor mixing matrix $\cal U$ in the (3+3) scenario
with three sterile neutrinos denoted as $\nu^{}_x$, $\nu^{}_y$
and $\nu^{}_z$:
\begin{eqnarray}
\left( \matrix{
\nu^{}_e \cr \nu^{}_\mu \cr \nu^{}_\tau \cr \nu^{}_x \cr \nu^{}_y
\cr \nu^{}_z \cr} \right) = {\cal U} \left( \matrix{
\nu^{}_1 \cr \nu^{}_2 \cr \nu^{}_3 \cr \nu^{}_4 \cr \nu^{}_5
\cr \nu^{}_6 \cr} \right) \; ,
\end{eqnarray}
where $\nu^{}_i$ (for $i=1, \cdots, 6$) stand for the mass eigenstates
of active and sterile neutrinos. Such a complete parametrization,
which has been lacking in the literature \cite{BF}, is expected to be
very useful for the study of neutrino phenomenology
at both low and high energy scales. We propose a simple but novel way to
establish the connection between active and sterile neutrinos in
terms of fifteen mixing angles and fifteen CP-violating phases. It
allows us to clearly describe the salient features of some interesting
problems, such as (a) possible contributions of light sterile neutrinos
to the tritium beta ($\beta$) decay and neutrinoless double-beta
($0\nu 2\beta$) decay; (b) leptonic CP violation
and deformed unitarity triangles of the $3\times 3$ flavor
mixing matrix of three active neutrinos; (c) a reconstruction of the
$6\times 6$ neutrino mass matrix in the type-(I+II) seesaw mechanism;
and (d) flavored and unflavored leptogenesis scenarios in the type-I
seesaw mechanism with three heavy Majorana neutrinos.

\section{The standard parametrization}

The $6\times 6$ unitary matrix $\cal U$ defined in Eq. (1) can be
decomposed as
\begin{eqnarray}
{\cal U} = \left( \matrix{
{\bf 1} & {\bf 0} \cr {\bf 0} & U^{}_0 \cr} \right)
\left( \matrix{A & R \cr S & B \cr} \right)
\left( \matrix{V^{}_0 & {\bf 0} \cr {\bf 0} & {\bf 1} \cr} \right) \; ,
\end{eqnarray}
in which ${\bf 0}$ and ${\bf 1}$ stand respectively for the $3\times 3$
zero and identity matrices, $U^{}_0$ and $V^{}_0$ are the $3\times 3$
unitary matrices, and $A$, $B$, $R$ and $S$ are the $3\times 3$
matrices which satisfy the conditions
\begin{eqnarray}
A A^\dagger + R R^\dagger = B B^\dagger + S S^\dagger = {\bf 1} \; ,
\nonumber \\
A S^\dagger + R B^\dagger = A^\dagger R + S^\dagger B = {\bf 0} \; ,
\nonumber \\
A^\dagger A + S^\dagger S = B^\dagger B + R^\dagger R = {\bf 1} \; ,
\end{eqnarray}
as a result of the unitarity of $\cal U$. In the limit of $R = S =
{\bf 0}$, $A = B = {\bf 1}$ holds and thus there is no correlation
between the active sector (described by $V^{}_0$) and the sterile
sector (characterized by $U^{}_0$). In view of Eq. (A7) in Appendix
A, we parametrize $\cal U$ as follows:
\begin{eqnarray}
\left( \matrix{V^{}_0 & {\bf 0} \cr {\bf 0} & {\bf 1} \cr} \right) &
= & O^{}_{23} O^{}_{13} O^{}_{12} \; ,
\nonumber \\
\left( \matrix{ {\bf 1} & {\bf 0} \cr {\bf 0} & U^{}_0 \cr} \right)
& = & O^{}_{56} O^{}_{46} O^{}_{45} \; ,
\nonumber \\
\left( \matrix{A & R \cr S & B \cr} \right) & = & O^{}_{36}
O^{}_{26} O^{}_{16} O^{}_{35} O^{}_{25} O^{}_{15} O^{}_{34}
O^{}_{24} O^{}_{14} \; ,
\end{eqnarray}
where fifteen two-dimensional rotation matrices $O^{}_{ij}$ (for
$1\leq i < j \leq 6$) in a six-dimensional complex space have been
given in Eqs. (A2)---(A6). To be explicit,
\begin{eqnarray}
V^{}_0 & = & \left( \matrix{ c^{}_{12} c^{}_{13} & \hat{s}^*_{12}
c^{}_{13} & \hat{s}^*_{13} \cr -\hat{s}^{}_{12} c^{}_{23} -
c^{}_{12} \hat{s}^{}_{13} \hat{s}^*_{23} & c^{}_{12} c^{}_{23} -
\hat{s}^*_{12} \hat{s}^{}_{13} \hat{s}^*_{23} & c^{}_{13}
\hat{s}^*_{23} \cr \hat{s}^{}_{12} \hat{s}^{}_{23} - c^{}_{12}
\hat{s}^{}_{13} c^{}_{23} & -c^{}_{12} \hat{s}^{}_{23} -
\hat{s}^*_{12} \hat{s}^{}_{13} c^{}_{23} & c^{}_{13} c^{}_{23} \cr}
\right) \; ,
\nonumber \\
U^{}_0 & = & \left( \matrix{ c^{}_{45} c^{}_{46} & \hat{s}^*_{45}
c^{}_{46} & \hat{s}^*_{46} \cr -\hat{s}^{}_{45} c^{}_{56} -
c^{}_{45} \hat{s}^{}_{46} \hat{s}^*_{56} & c^{}_{45} c^{}_{56} -
\hat{s}^*_{45} \hat{s}^{}_{46} \hat{s}^*_{56} & c^{}_{46}
\hat{s}^*_{56} \cr \hat{s}^{}_{45} \hat{s}^{}_{56} - c^{}_{45}
\hat{s}^{}_{46} c^{}_{56} & -c^{}_{45} \hat{s}^{}_{56} -
\hat{s}^*_{45} \hat{s}^{}_{46} c^{}_{56} & c^{}_{46} c^{}_{56} \cr}
\right) \; ,
\end{eqnarray}
in which $c^{}_{ij} \equiv \cos\theta^{}_{ij}$ and $\hat{s}^{}_{ij}
\equiv e^{i\delta^{}_{ij}} \sin\theta^{}_{ij}$ with $\theta^{}_{ij}$
and $\delta^{}_{ij}$ being the rotation angle and phase angle,
respectively. Both $V^{}_0$ and $U^{}_0$ have the standard form as
advocated in Ref. \cite{PDG}, and either of them consists of three
mixing angles and three CP-violating phases. If the sterile sector
is switched off, we are then left with the $3\times 3$ unitary
matrix $V^{}_0$ which describes the flavor mixing of three active
neutrinos. If the active sector is switched off, one will arrive at
the $3\times 3$ unitary matrix $U^{}_0$ which purely describes the
flavor mixing of three sterile neutrinos. In the type-I seesaw
mechanism \cite{SS}, for example, $U^{}_0$ is essentially equivalent
to the unitary transformation used to diagonalize the $3\times 3$
heavy Majorana neutrino mass matrix $M^{}_{\rm R}$ and therefore
relevant to the leptogenesis mechanism \cite{FY}.

With the help of Eq. (4) and Eqs. (A2)---(A6), a lengthy but
straightforward calculation leads us to the explicit expressions of
$A$, $B$, $R$ and $S$ as follows:
\begin{eqnarray}
A & = & \left( \matrix{ c^{}_{14} c^{}_{15} c^{}_{16} & 0 & 0
\cr\cr
\begin{array}{l} -c^{}_{14} c^{}_{15} \hat{s}^{}_{16} \hat{s}^*_{26} -
c^{}_{14} \hat{s}^{}_{15} \hat{s}^*_{25} c^{}_{26} \\
-\hat{s}^{}_{14} \hat{s}^*_{24} c^{}_{25} c^{}_{26} \end{array} &
c^{}_{24} c^{}_{25} c^{}_{26} & 0 \cr\cr
\begin{array}{l} -c^{}_{14} c^{}_{15} \hat{s}^{}_{16} c^{}_{26} \hat{s}^*_{36}
+ c^{}_{14} \hat{s}^{}_{15} \hat{s}^*_{25} \hat{s}^{}_{26} \hat{s}^*_{36} \\
- c^{}_{14} \hat{s}^{}_{15} c^{}_{25} \hat{s}^*_{35} c^{}_{36} +
\hat{s}^{}_{14} \hat{s}^*_{24} c^{}_{25} \hat{s}^{}_{26}
\hat{s}^*_{36} \\
+ \hat{s}^{}_{14} \hat{s}^*_{24} \hat{s}^{}_{25} \hat{s}^*_{35}
c^{}_{36} - \hat{s}^{}_{14} c^{}_{24} \hat{s}^*_{34} c^{}_{35}
c^{}_{36} \end{array} &
\begin{array}{l} -c^{}_{24} c^{}_{25} \hat{s}^{}_{26} \hat{s}^*_{36} -
c^{}_{24} \hat{s}^{}_{25} \hat{s}^*_{35} c^{}_{36} \\
-\hat{s}^{}_{24} \hat{s}^*_{34} c^{}_{35} c^{}_{36} \end{array} &
c^{}_{34} c^{}_{35} c^{}_{36} \cr} \right) \; ,
\nonumber \\
B & = & \left( \matrix{ c^{}_{14} c^{}_{24} c^{}_{34} & 0 & 0
\cr\cr
\begin{array}{l} -c^{}_{14} c^{}_{24} \hat{s}^{*}_{34} \hat{s}^{}_{35} -
c^{}_{14} \hat{s}^{*}_{24} \hat{s}^{}_{25} c^{}_{35} \\
-\hat{s}^{*}_{14} \hat{s}^{}_{15} c^{}_{25} c^{}_{35} \end{array} &
c^{}_{15} c^{}_{25} c^{}_{35} & 0 \cr\cr
\begin{array}{l} -c^{}_{14} c^{}_{24} \hat{s}^{*}_{34} c^{}_{35} \hat{s}^{}_{36}
+ c^{}_{14} \hat{s}^{*}_{24} \hat{s}^{}_{25} \hat{s}^{*}_{35} \hat{s}^{}_{36} \\
- c^{}_{14} \hat{s}^{*}_{24} c^{}_{25} \hat{s}^{}_{26} c^{}_{36} +
\hat{s}^{*}_{14} \hat{s}^{}_{15} c^{}_{25} \hat{s}^{*}_{35}
\hat{s}^{}_{36} \\
+ \hat{s}^{*}_{14} \hat{s}^{}_{15} \hat{s}^{*}_{25} \hat{s}^{}_{26}
c^{}_{36} - \hat{s}^{*}_{14} c^{}_{15} \hat{s}^{}_{16} c^{}_{26}
c^{}_{36} \end{array} &
\begin{array}{l} -c^{}_{15} c^{}_{25} \hat{s}^{*}_{35} \hat{s}^{}_{36} -
c^{}_{15} \hat{s}^{*}_{25} \hat{s}^{}_{26} c^{}_{36} \\
-\hat{s}^{*}_{15} \hat{s}^{}_{16} c^{}_{26} c^{}_{36} \end{array} &
c^{}_{16} c^{}_{26} c^{}_{36} \cr} \right) \; ;
\end{eqnarray}
and
\begin{eqnarray}
R & = & \left( \matrix{ \hat{s}^*_{14} c^{}_{15} c^{}_{16} &
\hat{s}^*_{15} c^{}_{16} & \hat{s}^*_{16} \cr\cr
\begin{array}{l} -\hat{s}^*_{14} c^{}_{15} \hat{s}^{}_{16} \hat{s}^*_{26} -
\hat{s}^*_{14} \hat{s}^{}_{15} \hat{s}^*_{25} c^{}_{26} \\
+ c^{}_{14} \hat{s}^*_{24} c^{}_{25} c^{}_{26} \end{array} & -
\hat{s}^*_{15} \hat{s}^{}_{16} \hat{s}^*_{26} + c^{}_{15}
\hat{s}^*_{25} c^{}_{26} & c^{}_{16} \hat{s}^*_{26} \cr\cr
\begin{array}{l} -\hat{s}^*_{14} c^{}_{15} \hat{s}^{}_{16} c^{}_{26}
\hat{s}^*_{36} + \hat{s}^*_{14} \hat{s}^{}_{15} \hat{s}^*_{25}
\hat{s}^{}_{26} \hat{s}^*_{36} \\ - \hat{s}^*_{14} \hat{s}^{}_{15}
c^{}_{25} \hat{s}^*_{35} c^{}_{36} - c^{}_{14} \hat{s}^*_{24}
c^{}_{25} \hat{s}^{}_{26}
\hat{s}^*_{36} \\
- c^{}_{14} \hat{s}^*_{24} \hat{s}^{}_{25} \hat{s}^*_{35}
c^{}_{36} + c^{}_{14} c^{}_{24} \hat{s}^*_{34} c^{}_{35} c^{}_{36}
\end{array} &
\begin{array}{l} -\hat{s}^*_{15} \hat{s}^{}_{16} c^{}_{26} \hat{s}^*_{36}
- c^{}_{15} \hat{s}^*_{25} \hat{s}^{}_{26} \hat{s}^*_{36} \\
+c^{}_{15} c^{}_{25} \hat{s}^*_{35} c^{}_{36} \end{array} &
c^{}_{16} c^{}_{26} \hat{s}^*_{36} \cr} \right) \; ,
\nonumber \\
S & = & \left( \matrix{ -\hat{s}^{}_{14} c^{}_{24} c^{}_{34} &
-\hat{s}^{}_{24} c^{}_{34} & -\hat{s}^{}_{34} \cr\cr
\begin{array}{l} \hat{s}^{}_{14} c^{}_{24} \hat{s}^{*}_{34} \hat{s}^{}_{35}
+ \hat{s}^{}_{14} \hat{s}^{*}_{24} \hat{s}^{}_{25} c^{}_{35} \\
- c^{}_{14} \hat{s}^{}_{15} c^{}_{25} c^{}_{35} \end{array} &
\hat{s}^{}_{24} \hat{s}^{*}_{34} \hat{s}^{}_{35} - c^{}_{24}
\hat{s}^{}_{25} c^{}_{35} & -c^{}_{34} \hat{s}^{}_{35} \cr\cr
\begin{array}{l} \hat{s}^{}_{14} c^{}_{24} \hat{s}^{*}_{34} c^{}_{35}
\hat{s}^{}_{36} - \hat{s}^{}_{14} \hat{s}^{*}_{24} \hat{s}^{}_{25}
\hat{s}^{*}_{35} \hat{s}^{}_{36} \\ + \hat{s}^{}_{14}
\hat{s}^{*}_{24} c^{}_{25} \hat{s}^{}_{26} c^{}_{36} + c^{}_{14}
\hat{s}^{}_{15} c^{}_{25} \hat{s}^{*}_{35}
\hat{s}^{}_{36} \\
+ c^{}_{14} \hat{s}^{}_{15} \hat{s}^{*}_{25} \hat{s}^{}_{26}
c^{}_{36} - c^{}_{14} c^{}_{15} \hat{s}^{}_{16} c^{}_{26} c^{}_{36}
\end{array} &
\begin{array}{l} \hat{s}^{}_{24} \hat{s}^{*}_{34} c^{}_{35} \hat{s}^{}_{36}
+ c^{}_{24} \hat{s}^{}_{25} \hat{s}^{*}_{35} \hat{s}^{}_{36} \\
-c^{}_{24} c^{}_{25} \hat{s}^{}_{26} c^{}_{36} \end{array} &
-c^{}_{34} c^{}_{35} \hat{s}^{}_{36} \cr} \right) \; .
\end{eqnarray}
We see that the textures of $A$ and $B$ are rather similar, so are
the textures of $R$ and $S$. In fact, the expression of $B$ can be
obtained from that of $A^*$ with the subscript replacements $15
\leftrightarrow 24$, $16 \leftrightarrow 34$, and $26
\leftrightarrow 35$; and the expression of $S$ can be obtained from
that of $-R^*$ with the same subscript replacements. Note that the
results of $A$ and $R$ have been obtained in Ref. \cite{Xing08},
and here we present the results of $B$
and $S$ to complete a full parametrization of the $6\times 6$ flavor
mixing matrix $\cal U$.

It proves convenient to define $V \equiv A V^{}_0$ and $U \equiv
U^{}_0 B$ which describe the flavor mixing phenomena of three active
neutrinos and three sterile neutrinos, respectively. Furthermore,
$\widehat{S} \equiv U^{}_0 S V^{}_0$ links the mass eigenstates
$(\nu^{}_1, \nu^{}_2, \nu^{}_3)$ to the sterile flavor eigenstates
$(\nu^{}_x, \nu^{}_y, \nu^{}_z)$ in the chosen basis. We therefore
have
\begin{eqnarray}
\left(\matrix{\nu^{}_e \cr \nu^{}_\mu \cr \nu^{}_\tau \cr} \right) =
V \left(\matrix{\nu^{}_1 \cr \nu^{}_2 \cr \nu^{}_3 \cr} \right) + R
\left(\matrix{\nu^{}_4 \cr \nu^{}_5 \cr \nu^{}_6 \cr} \right) \; ,
\end{eqnarray}
and
\begin{eqnarray}
\left(\matrix{\nu^{}_x \cr \nu^{}_y \cr \nu^{}_z \cr} \right) = U
\left(\matrix{\nu^{}_4 \cr \nu^{}_5 \cr \nu^{}_6 \cr} \right) \; +
\widehat{S} \left(\matrix{\nu^{}_1 \cr \nu^{}_2 \cr \nu^{}_3 \cr}
\right) \; .
\end{eqnarray}
Eq. (8) directly leads us to the standard weak charged-current
interactions of six neutrinos:
\begin{eqnarray}
-{\cal L}^{}_{\rm cc} = \frac{g}{\sqrt{2}} ~ \overline{\left(e~~
\mu~~ \tau\right)^{}_{\rm L}} ~ \gamma^\mu \left[ V \left(
\matrix{\nu^{}_1 \cr \nu^{}_2 \cr \nu^{}_3} \right)^{}_{\rm L} + R
\left( \matrix{\nu^{}_4 \cr \nu^{}_5 \cr \nu^{}_6} \right)^{}_{\rm
L} \right] W^-_\mu + {\rm h.c.} \; ,
\end{eqnarray}
where $V$ is just the Maki-Nakagawa-Sakata-Pontecorvo (MNSP) matrix
\cite{MNS} responsible for the active neutrino mixing, and $R$
measures the strength of charged-current interactions between $(e,
\mu, \tau)$ and $(\nu^{}_4, \nu^{}_5, \nu^{}_6)$. Because of
\begin{eqnarray}
VV^\dagger = AA^\dagger = {\bf 1} - RR^\dagger \; ,
\nonumber \\
U^\dagger U = B^\dagger B = {\bf 1} - R^\dagger R \; ,
\end{eqnarray}
we find that both $V$ and $U$ are not exactly unitary and their
non-unitary effects are simply characterized by non-vanishing $R$
and $S$.

In view of current observational constraints on sterile neutrinos,
we expect that the mixing angles between active and sterile
neutrinos are strongly suppressed (at most at the ${\cal O}(0.1)$ level
\cite{Schwetz}
\footnote{For example, the non-unitarity of $V = A V^{}_0$ or the
deviation of $V$ from $V^{}_0$ can at most be at the $1\%$ level as
constrained by current neutrino oscillation data and precision
electroweak data \cite{Antusch}.}).
The smallness of $\theta^{}_{ij}$ (for $i=1,2,3$ and $j=4,5,6$)
allows us to make the following excellent approximations to
Eqs. (7) and (8):
\begin{eqnarray}
A & \simeq & {\bf 1} - \left( \matrix{ \frac{1}{2} \left( s^2_{14} +
s^2_{15} + s^2_{16} \right) & 0 & 0 \cr \hat{s}^{}_{14}
\hat{s}^*_{24} + \hat{s}^{}_{15} \hat{s}^*_{25} + \hat{s}^{}_{16}
\hat{s}^*_{26} & \frac{1}{2} \left( s^2_{24} + s^2_{25} + s^2_{26}
\right) & 0 \cr \hat{s}^{}_{14} \hat{s}^*_{34} + \hat{s}^{}_{15}
\hat{s}^*_{35} + \hat{s}^{}_{16} \hat{s}^*_{36} & \hat{s}^{}_{24}
\hat{s}^*_{34} + \hat{s}^{}_{25} \hat{s}^*_{35} + \hat{s}^{}_{26}
\hat{s}^*_{36} & \frac{1}{2} \left( s^2_{34} +
s^2_{35} + s^2_{36} \right) \cr} \right) \; ,
\nonumber \\
B & \simeq & {\bf 1} - \left( \matrix{ \frac{1}{2} \left( s^2_{14} +
s^2_{24} + s^2_{34} \right) & 0 & 0 \cr \hat{s}^{*}_{14}
\hat{s}^{}_{15} + \hat{s}^{*}_{24} \hat{s}^{}_{25} + \hat{s}^{*}_{34}
\hat{s}^{}_{35} & \frac{1}{2} \left( s^2_{15} + s^2_{25} + s^2_{35}
\right) & 0 \cr \hat{s}^{*}_{14} \hat{s}^{}_{16} + \hat{s}^{*}_{24}
\hat{s}^{}_{26} + \hat{s}^{*}_{34} \hat{s}^{}_{36} & \hat{s}^{*}_{15}
\hat{s}^{}_{16} + \hat{s}^{*}_{25} \hat{s}^{}_{26} + \hat{s}^{*}_{35}
\hat{s}^{}_{36} & \frac{1}{2} \left( s^2_{16} +
s^2_{26} + s^2_{36} \right) \cr} \right) \; ,
\end{eqnarray}
where the terms of ${\cal O}(s^4_{ij})$ have been omitted; and
\begin{eqnarray}
R & \simeq & {\bf 0} + \left( \matrix{ \hat{s}^*_{14} & \hat{s}^*_{15}
& \hat{s}^*_{16} \cr \hat{s}^*_{24} & \hat{s}^*_{25} &
\hat{s}^*_{26} \cr \hat{s}^*_{34} & \hat{s}^*_{35} &
\hat{s}^*_{36} \cr} \right) \; ,
\nonumber \\
S & \simeq & {\bf 0} - \left( \matrix{ \hat{s}^{}_{14} & \hat{s}^{}_{24}
& \hat{s}^{}_{34} \cr \hat{s}^{}_{15} & \hat{s}^{}_{25} &
\hat{s}^{}_{35} \cr \hat{s}^{}_{16} & \hat{s}^{}_{26} &
\hat{s}^{}_{36} \cr} \right) \; ,
\end{eqnarray}
where the terms of ${\cal O}(s^3_{ij})$ have been omitted. It turns
out that $R \simeq -S^\dagger$ holds in the same approximation.

Note that the $6 \times 6$ unitary matrix $\cal U$ can be used
to describe not only the flavor mixing between active and sterile
neutrinos but also the flavor mixing between ordinary and extra quarks.
Note also that it is straightforward to obtain the (3+1) flavor
mixing scenario from Eqs. (6) and (7) by switching off the
mixing angles $\theta^{}_{i5}$ and $\theta^{}_{j6}$ (for
$1 \leq i \leq 4$ and $1 \leq j \leq 5$), or the (3+2) flavor mixing
scenario by turning off the mixing angles $\theta^{}_{j6}$ (for
$1 \leq j \leq 5$).

\section{Some applications}

To illustrate the usefulness of our parametrization of the
$6\times 6$ flavor mixing matrix $\cal U$, let us briefly discuss
its four simple but instructive applications in
neutrino phenomenology.

\subsection{The effective masses of $\beta$ and $0\nu 2\beta$ decays}

One or two light sterile neutrinos at the sub-eV
mass scale have been hypothesized for a quite long time to interpret
the LSND antineutrino anomaly \cite{LSND} and the subsequent
MiniBooNE antineutrino puzzle \cite{M}. In general, however,
it seems more natural to assume the number of sterile neutrino
species to be equal to that of active neutrino species \cite{MS}
\footnote{In order to avoid any severe conflict between such a (3+3)
scenario and the standard $\Lambda$CDM cosmology, it is perhaps necessary
to either loosen the mass hierarchy of three sterile neutrinos
(i.e., not all of them are of ${\cal O}(0.1)$ eV) or refer to
some nonstandard models of cosmology \cite{Raffelt}.},
such that even possible warm dark matter in the form of one or two species
of keV sterile neutrinos could be taken into account.

For simplicity and illustration, we are only concerned about the effective
masses of the tritium beta ($\beta$) decay $^3_1{\rm H} \to ~^3_2{\rm He}
+ e^- + \overline{\nu}^{}_e$ and the neutrinoless
double-beta ($0\nu 2\beta$) decay $A(Z,A) \to A(Z+2, N-2) + 2 e^-$
in the (3+3) neutrino mixing scenario. The former is
\begin{eqnarray}
\langle m\rangle^{\prime}_{e} & \equiv & \left[ \sum_{i=1}^6 m^2_i
|V^{}_{e i}|^2 \right]^{1/2} =
\sqrt{\langle m\rangle^2_e c^2_{14} c^2_{15} c^2_{16} +
m^2_4 s^2_{14} c^2_{15} c^2_{16} + m^2_5 s^2_{15} c^2_{16} +
m^2_6 s^2_{16}} \;\; ,
\end{eqnarray}
where $\langle m\rangle^{}_e = \sqrt{m^2_1 c^2_{12} c^2_{13}
+ m^2_2 s^2_{12} c^2_{13} + m^2_3 s^2_{13}}$ is the standard
contribution from three active neutrinos. We see that
$\langle m\rangle^\prime_e \geq \langle m\rangle^{}_e$ always
holds. The effective mass of the $0\nu 2\beta$ decay is
\begin{eqnarray}
\langle m\rangle^{\prime}_{ee} & \equiv & \sum_{i=1}^6 m^{}_i
V^2_{e i}
= \langle m\rangle^{}_{ee} \left(c^{}_{14} c^{}_{15} c^{}_{16}
\right)^2 + m^{}_4 \left(\hat{s}^*_{14} c^{}_{15} c^{}_{16}\right)^2 +
m^{}_5 \left(\hat{s}^*_{15} c^{}_{16}\right)^2 + m^{}_6
\left(\hat{s}^*_{16}\right)^2 \;
\end{eqnarray}
with $\langle m\rangle^{}_{ee} = m^{}_1 (c^{}_{12} c^{}_{13})^2 +
m^{}_2 (\hat{s}^*_{12} c^{}_{13})^2 + m^{}_3 (\hat{s}^*_{13})^2$
being the standard contribution from three active neutrinos. It is
difficult to say about the relative magnitudes of $\langle
m\rangle^{}_{ee}$ and $\langle m\rangle^{\prime}_{ee}$, because the
CP-violating phases may give rise to more or less cancelations of
different terms in them. In particular, even $\langle
m\rangle^{}_{ee} =0$ \cite{Xing03} or $\langle
m\rangle^{\prime}_{ee} =0$ \cite{LL} is not impossible. If both
$\langle m\rangle^\prime_e$ and $\langle m\rangle^{\prime}_{ee}$ can
be determined or constrained in the future experiments, a comparison
between them might be able to probe the existence of light sterile
neutrinos \cite{Rodejohann}.

\subsection{Deformed unitarity triangles and CP violation}

Switching off three sterile neutrinos, one may describe flavor
mixing and CP violation of three active neutrinos in terms of six
unitarity triangles in the complex plane \cite{FX99}. Three of them,
defined by the orthogonality conditions
\begin{eqnarray}
\triangle^{}_e : &~~& V^{}_{\mu 1} V^*_{\tau 1} + V^{}_{\mu 2}
V^*_{\tau 2} + V^{}_{\mu 3} V^*_{\tau 3} = 0 \; , \nonumber \\
\triangle^{}_\mu : && V^{}_{\tau 1} V^*_{e 1} + V^{}_{\tau 2}
V^*_{e 2} + V^{}_{\tau 3} V^*_{e 3} = 0 \; , \nonumber \\
\triangle^{}_\tau : && V^{}_{e 1} V^*_{\mu 1} + V^{}_{e 2} V^*_{\mu
2} + V^{}_{e 3} V^*_{\mu 3} = 0 \; ,
\end{eqnarray}
are illustrated in FIG. 1 (left panel). The area of each triangle
is equal to $J^{}_0/2$, where $J^{}_0$ is the Jarlskog parameter
given in Eq. (B2) and measures the strength of leptonic CP-violating
effects in $\nu^{}_\mu \to \nu^{}_\tau$, $\nu^{}_\tau \to \nu^{}_e$
and $\nu^{}_e \to \nu^{}_\mu$ oscillations. Now let us turn on the
contributions of three sterile neutrinos to flavor mixing and CP
violation. Then Eq. (16) approximates to
\begin{eqnarray}
\triangle^{\prime}_e : &~~& V^{}_{\mu 1} V^*_{\tau 1} + V^{}_{\mu 2}
V^*_{\tau 2} + V^{}_{\mu 3} V^*_{\tau 3} \simeq -{\cal Z}^* \; , \nonumber \\
\triangle^{\prime}_\mu : && V^{}_{\tau 1} V^*_{e 1} + V^{}_{\tau 2}
V^*_{e 2} + V^{}_{\tau 3} V^*_{e 3} \simeq -{\cal Y} \; , \nonumber \\
\triangle^{\prime}_\tau : && V^{}_{e 1} V^*_{\mu 1} + V^{}_{e 2}
V^*_{\mu 2} + V^{}_{e 3} V^*_{\mu 3} \simeq -{\cal X}^* \; ,
\end{eqnarray}
where ${\cal X} \equiv \hat{s}^{}_{14} \hat{s}^*_{24} +
\hat{s}^{}_{15} \hat{s}^*_{25} + \hat{s}^{}_{16} \hat{s}^*_{26}$,
${\cal Y} \equiv \hat{s}^{}_{14} \hat{s}^*_{34} + \hat{s}^{}_{15}
\hat{s}^*_{35} + \hat{s}^{}_{16} \hat{s}^*_{36}$ and ${\cal Z}
\equiv \hat{s}^{}_{24} \hat{s}^*_{34} + \hat{s}^{}_{25}
\hat{s}^*_{35} + \hat{s}^{}_{26} \hat{s}^*_{36}$. These deformed
unitarity triangles are also illustrated in FIG. 1 (right panel).
The small differences of their areas from $J^{}_0/2$ just signify
the new CP-violating effects.

Let us take a look at the CP-violating asymmetries between
$\nu^{}_\alpha \to \nu^{}_\beta$ and $\overline{\nu}^{}_\alpha \to
\overline{\nu}^{}_\beta$ oscillations, defined as ${\cal
A}^{}_{\alpha\beta} \equiv P(\nu^{}_\alpha \to \nu^{}_\beta) -
P(\overline{\nu}^{}_\alpha \to \overline{\nu}^{}_\beta)$. With the
help of Eq. (B5), we explicitly obtain
\begin{eqnarray}
{\cal A}^{}_{\mu e} & \simeq & -4 \left( J^{}_0 + c^{}_{12}
s^{}_{12} c^{}_{23} {\rm Im} X \right) \sin\frac{\Delta m^2_{21}
L}{2E} \; , \nonumber \\
{\cal A}^{}_{e\tau} & \simeq & -4 \left( J^{}_0 + c^{}_{12}
s^{}_{12} s^{}_{23} {\rm Im} Y \right) \sin\frac{\Delta m^2_{21}
L}{2E} \; ,
\nonumber \\
{\cal A}^{}_{\mu\tau} & \simeq & +4 \left[ J^{}_0 + c^{}_{12}
s^{}_{12} c^{}_{23} s^{}_{23} \left( s^{}_{23} {\rm Im} X +
c^{}_{23} {\rm Im} Y \right) \right] \sin\frac{\Delta m^2_{21}
L}{2E} + 4 c^{}_{23} s^{}_{23} {\rm Im} Z \sin\frac{\Delta m^2_{32}
L}{2E} \; ,
\end{eqnarray}
where $X \equiv {\cal X} e^{-i\delta^{}_{12}}$, $Y \equiv {\cal Y}
e^{-i (\delta^{}_{12} + \delta^{}_{23})}$ and $Z \equiv {\cal Z}
e^{-i\delta^{}_{23}}$. We see that ${\cal A}^{}_{\mu e}$ and ${\cal
A}^{}_{e \tau}$ are related to the deformed unitarity triangles
$\triangle^\prime_\tau$ and $\triangle^\prime_\mu$, respectively. In
comparison, $\triangle^\prime_e$ has something to do with ${\cal
A}^{}_{\mu \tau}$. It is therefore possible to determine three new
CP-violating terms ${\rm Im} X$, ${\rm Im} Y$ and ${\rm Im} Z$ by
measuring the CP-violating effects in neutrino oscillations. Note
that three CP-violating asymmetries in Eq. (18) satisfy the
correlation
\begin{eqnarray}
{\cal A}^{}_{\mu \tau} + \left( s^2_{23} {\cal A}^{}_{\mu e} +
c^2_{23} {\cal A}^{}_{e \tau} \right) \simeq 4 c^{}_{23} s^{}_{23}
{\rm Im} Z \sin\frac{\Delta m^2_{32} L}{2E} \; .
\end{eqnarray}
When $\Delta m^2_{32} L/E \sim \pi$ holds, both ${\cal A}^{}_{\mu
e}$ and ${\cal A}^{}_{e \tau}$ are suppressed such that ${\cal
A}^{}_{\mu \tau}$ becomes a pure measure of the non-unitary
CP-violating parameter ${\rm Im} Z$. This interesting possibility,
together with terrestrial matter effects, has been discussed before
(e.g., Refs. \cite{Xing08} and \cite{Non}).

\subsection{Reconstruction of the $6\times 6$ neutrino mass matrix}

The type-(I+II) seesaw mechanism \cite{T2} is a good example to
illustrate the flavor mixing between three active neutrinos and
three heavy Majorana neutrinos. In this mechanism the mass term of
six neutrinos is usually written as
\begin{eqnarray}
-{\cal L}^{}_{\rm mass} = \frac{1}{2} ~ \overline{\left( \nu^{}_{\rm
L} ~N^c_{\rm R}\right)} ~ \left( \matrix{ M^{}_{\rm L} & M^{}_{\rm D} \cr
M^T_{\rm D} & M^{}_{\rm R}}\right) \left( \matrix{ \nu^c_{\rm L} \cr
N^{}_{\rm R}}\right) + {\rm h.c.} \; ,
\end{eqnarray}
where $\nu^{}_{\rm L}$ and $N^{}_{\rm R}$ represent the column
vectors of three left-handed neutrinos and three right-handed
neutrinos, respectively. The overall $6\times 6$ neutrino mass
matrix in Eq. (20) can be diagonalized by a unitary transformation:
\begin{eqnarray}
{\cal U}^\dagger \left( \matrix{ M^{}_{\rm L} & M^{}_{\rm D} \cr M^T_{\rm
D} & M^{}_{\rm R}}\right) {\cal U}^* = \left( \matrix{
\widehat{M}^{}_\nu & {\bf 0} \cr {\bf 0} & \widehat{M}^{}_N}\right)
\; ,
\end{eqnarray}
where $\cal U$ is already given in Eq. (2), $\widehat{M}^{}_\nu
\equiv {\rm Diag}\{m^{}_1, m^{}_2, m^{}_3 \}$ and $\widehat{M}^{}_N
\equiv {\rm Diag}\{M^{}_1, M^{}_2, M^{}_3 \}$ with $m^{}_i$ or
$M^{}_i$ (for $i=1,2,3$) being the physical masses of light or heavy
Majorana neutrinos. The standard weak charged-current interactions
of six neutrinos are given by Eq. (10) with $\nu^{}_4 = M^{}_1$,
$\nu^{}_5 = N^{}_2$ and $\nu^{}_6 = N^{}_3$ in the basis of mass
eigenstates. With the help of Eq. (2),
\begin{eqnarray}
M^{}_{\rm L} & = & V \widehat{M}^{}_\nu V^T + R \widehat{M}^{}_N R^T
\simeq V^{}_0 \widehat{M}^{}_\nu V^T_0 + R \widehat{M}^{}_N R^T \; ,
\nonumber \\
M^{}_{\rm D} & = & V \widehat{M}^{}_\nu \widehat{S}^T +
R \widehat{M}^{}_N U^T \simeq R \widehat{M}^{}_N U^T_0 \; ,
\nonumber \\
M^{}_{\rm R} & = & \widehat{S} \widehat{M}^{}_\nu \widehat{S}^T
+ U \widehat{M}^{}_N U^T \simeq U^{}_0 \widehat{M}^{}_N U^T_0 \; ,
\end{eqnarray}
where $V\equiv A V^{}_0$, $U \equiv U^{}_0 B$ and
$\widehat{S} \equiv U^{}_0 S V^{}_0$ have been defined before. The
approximations made on the right-hand side of Eq. (22) follow the
spirit that only the leading terms of $M^{}_{\rm L}$, $M^{}_{\rm D}$
and $M^{}_{\rm R}$ are kept. It is then possible to reconstruct
these $3\times 3$ neutrino mass matrices, at least in principle,
in terms of neutrino masses and flavor mixing parameters \cite{Zralek}.

Given the basis where $M^{}_{\rm R}$ is diagonal, real and positive,
Eq. (22) implies that $M^{}_{\rm R} \simeq \widehat{M}^{}_N$ together
with $U^{}_0 \simeq {\bf 1}$ is a good approximation. Note that such
a flavor basis is often chosen in the study of leptogenesis, because
the decays of $N^{}_i$ (for $i=1,2,3$) need to be calculated. It is
also much easier to reconstruct $M^{}_{\rm D}$ and $M^{}_{\rm L}$
in this special basis. For example,
\begin{eqnarray}
M^{}_{\rm D} \simeq R \widehat{M}^{}_N \simeq \left( \matrix{
M^{}_1 \hat{s}^*_{14} & M^{}_2 \hat{s}^*_{15}
& M^{}_3 \hat{s}^*_{16} \cr M^{}_1 \hat{s}^*_{24} & M^{}_2 \hat{s}^*_{25} &
M^{}_3 \hat{s}^*_{26} \cr M^{}_1 \hat{s}^*_{34} & M^{}_2 \hat{s}^*_{35} &
M^{}_3 \hat{s}^*_{36} \cr} \right) \; ;
\end{eqnarray}
and six independent matrix elements of
$M^{}_{\rm L} \simeq V^{}_0 \widehat{M}^{}_\nu V^T_0 + M^{}_{\rm D} R^T$
can similarly be obtained:
\begin{eqnarray}
(M^{}_{\rm L})^{}_{ee} & \simeq & m^{}_1 \left(c^{}_{12} c^{}_{13} \right)^2
+ m^{}_2 \left( \hat{s}^*_{12} c^{}_{13} \right)^2 + m^{}_3 \left(
\hat{s}^*_{13} \right)^2 + M^{}_1 \left( \hat{s}^*_{14} \right)^2 +
M^{}_2 \left( \hat{s}^*_{15} \right)^2 + M^{}_3 \left( \hat{s}^*_{16} \right)^2
\; ,
\nonumber \\
(M^{}_{\rm L})^{}_{e\mu} & \simeq & -m^{}_1 c^{}_{12} c^{}_{13}
\left( \hat{s}^{}_{12} c^{}_{23} + c^{}_{12} \hat{s}^{}_{13} \hat{s}^*_{23}
\right) + m^{}_2 \hat{s}^*_{12} c^{}_{13}
\left( c^{}_{12} c^{}_{23} - \hat{s}^*_{12} \hat{s}^{}_{13} \hat{s}^{*}_{23}
\right) + m^{}_3 c^{}_{13} \hat{s}^*_{13} \hat{s}^{*}_{23}
\nonumber \\
&& + M^{}_1 \hat{s}^*_{14} \hat{s}^*_{24} +
M^{}_2 \hat{s}^*_{15} \hat{s}^*_{25} + M^{}_3 \hat{s}^*_{16} \hat{s}^*_{26}
\; ,
\nonumber \\
(M^{}_{\rm L})^{}_{e\tau} & \simeq & m^{}_1 c^{}_{12} c^{}_{13}
\left( \hat{s}^{}_{12} \hat{s}^{}_{23} - c^{}_{12} \hat{s}^{}_{13} c^{}_{23}
\right) - m^{}_2 \hat{s}^*_{12} c^{}_{13}
\left( c^{}_{12} \hat{s}^{}_{23} + \hat{s}^*_{12} \hat{s}^{}_{13} c^{}_{23}
\right) + m^{}_3 c^{}_{13} \hat{s}^*_{13} c^{}_{23}
\nonumber \\
&& + M^{}_1 \hat{s}^*_{14} \hat{s}^*_{34} +
M^{}_2 \hat{s}^*_{15} \hat{s}^*_{35} + M^{}_3 \hat{s}^*_{16} \hat{s}^*_{36}
\; ,
\nonumber \\
(M^{}_{\rm L})^{}_{\mu\mu} & \simeq & m^{}_1 \left( \hat{s}^{}_{12} c^{}_{23}
+ c^{}_{12} \hat{s}^{}_{13} \hat{s}^*_{23} \right)^2 + m^{}_2
\left( c^{}_{12} c^{}_{23} - \hat{s}^*_{12} \hat{s}^{}_{13} \hat{s}^*_{23}
\right)^2 + m^{}_3 \left( c^{}_{13} \hat{s}^*_{23} \right)^2
\nonumber \\
&& + M^{}_1 \left( \hat{s}^*_{24} \right)^2 +
M^{}_2 \left( \hat{s}^*_{25} \right)^2 + M^{}_3 \left( \hat{s}^*_{26} \right)^2
\; ,
\nonumber \\
(M^{}_{\rm L})^{}_{\mu\tau} & \simeq & -m^{}_1 \left(\hat{s}^{}_{12} c^{}_{23}
+ c^{}_{12} \hat{s}^{}_{13} \hat{s}^*_{23} \right)
\left( \hat{s}^{}_{12} \hat{s}^{}_{23} - c^{}_{12} \hat{s}^{}_{13} c^{}_{23}
\right)
\nonumber \\
&& - m^{}_2 \left( c^{}_{12} c^{}_{23} - \hat{s}^*_{12} \hat{s}^{}_{13}
\hat{s}^*_{23} \right)
\left( c^{}_{12} \hat{s}^{}_{23} + \hat{s}^*_{12} \hat{s}^{}_{13} c^{}_{23}
\right) + m^{}_3 c^2_{13} c^{}_{23} \hat{s}^*_{23}
\nonumber \\
&& + M^{}_1 \hat{s}^*_{24} \hat{s}^*_{34} +
M^{}_2 \hat{s}^*_{25} \hat{s}^*_{35} + M^{}_3 \hat{s}^*_{26} \hat{s}^*_{36}
\; ,
\nonumber \\
(M^{}_{\rm L})^{}_{\tau\tau} & \simeq & m^{}_1
\left( \hat{s}^{}_{12} \hat{s}^{}_{23}
- c^{}_{12} \hat{s}^{}_{13} c^{}_{23} \right)^2 + m^{}_2
\left( c^{}_{12} \hat{s}^{}_{23} + \hat{s}^*_{12} \hat{s}^{}_{13} c^{}_{23}
\right)^2 + m^{}_3 \left( c^{}_{13} c^{}_{23} \right)^2
\nonumber \\
&& + M^{}_1 \left( \hat{s}^*_{34} \right)^2 +
M^{}_2 \left( \hat{s}^*_{35} \right)^2 + M^{}_3 \left( \hat{s}^*_{36} \right)^2
\; .
\end{eqnarray}
So any specific textures of $M^{}_{\rm D}$ and
$M^{}_{\rm L}$ predicted in a specific type-(I+II) seesaw model must
have direct and important impacts on the magnitudes of neutrino masses,
flavor mixing angles and CP-violating phases.

\subsection{Flavored and unflavored leptogenesis scenarios}

It is straightforward to obtain the type-I seesaw mechanism from the
type-(I+II) seesaw mechanism by switching off the $M^{}_{\rm L}$ term.
In this special case one arrives at the exact type-I seesaw
relation $V \widehat{M}^{}_\nu V^T + R \widehat{M}^{}_N R^T = {\bf
0}$, where $V$ and $R$ satisfy the unitarity condition
$VV^\dagger + RR^\dagger = {\bf 1}$.
Given $M^{}_i \gg m^{}_i$, it is more popular to write the $3\times 3$
light Majorana neutrino mass matrix as
\begin{equation} M^{}_\nu \equiv V
\widehat{M}^{}_\nu V^T = -R \widehat{M}^{}_N R^T \simeq -M^{}_{\rm
D} M^{-1}_{\rm R} M^T_{\rm D} \; ,
\end{equation}
where $V \equiv A V^{}_0 \simeq V^{}_0$ holds in the same
approximation \cite{Xing10}. Associated with this seesaw picture,
the leptogenesis mechanism \cite{FY} provides a natural possibility
of accounting for the observed matter-antimatter asymmetry of the
Universe via the lepton-number-violating, CP-violating and
out-of-equilibrium decays of $N^{}_i$ and the $(B-L)$-conserving
sphaleron processes. The CP-violating asymmetry between
$N^{}_i \rightarrow \ell^{}_\alpha + H$ and
$N^{}_i \rightarrow \overline{\ell}^{}_\alpha + \overline{H}$ decays,
denoted as $\varepsilon^{}_{i\alpha}$ (for $i=1,2,3$ and
$\alpha = e, \mu, \tau$), is given by \cite{Book}
\begin{eqnarray}
\varepsilon^{}_{i\alpha} &\equiv& \frac{\Gamma(N^{}_i \to \ell^{}_\alpha
+ H) - \Gamma(N^{}_i \to \overline{\ell}^{}_\alpha
+ \overline{H})} {\Gamma(N^{}_i \to \ell^{}_\alpha
+ H) + \Gamma(N^{}_i \to \overline{\ell}^{}_\alpha
+ \overline{H})}
\nonumber \\
&=& \frac{1}{8\pi v^2 (M^\dagger_{\rm D}
M^{}_{\rm D})^{}_{ii}} \sum_{j \neq i} \left\{ {\rm
Im}\left[(M^*_{\rm D})^{}_{\alpha i} (M^{}_{\rm D})^{}_{\alpha j}
(M^\dagger_{\rm D} M^{}_{\rm D})^{}_{ij}\right] {\cal
F}\left(\frac{M^2_j}{M^2_i}\right) \right. \nonumber \\
&& ~~~~~~~~~~~~~~~~~~~~~~~ + \left. {\rm Im}\left[(M^*_{\rm D})^{}_{\alpha
i} (M^{}_{\rm D})^{}_{\alpha j} (M^\dagger_{\rm D} M^{}_{\rm D})^*_{ij}\right]
{\cal G}\left(\frac{M^2_j}{M^2_i}\right) \right\} \; ,
\end{eqnarray}
where the loop functions ${\cal F}(x) = \sqrt{x}
\{ (2-x)/(1-x) + (1+x) \ln [x/(1+x)] \}$ and ${\cal G}(x) = 1/(1-x)$
have been introduced. If all the interactions in the era of leptogenesis
were blind to lepton flavors, then only the total CP-violating asymmetry
$\varepsilon^{}_i$ should be relevant:
\begin{eqnarray}
\varepsilon^{}_i = \sum_\alpha \varepsilon^{}_{i\alpha} =
\frac{1}{8\pi (M^\dagger_{\rm D} M^{}_{\rm D})^{}_{ii}} \sum_{j \neq i} {\rm
Im} \left[(M^\dagger_{\rm D} M^{}_{\rm D})^2_{ij}\right] {\cal
F}\left(\frac{M^2_j}{M^2_i}\right) \; .
\end{eqnarray}
The leptogenesis mechanisms associated with Eqs. (26) and (27) are
usually referred to as flavored and unflavored leptogenesis scenarios,
respectively.

In the flavor basis where $M_{\rm R} \simeq \widehat{M}^{}_N$ and
$U^{}_0 \simeq {\bf 1}$ hold, we have obtained the explicit expression of
$M^{}_{\rm D} \simeq R \widehat{M}^{}_N$ in Eq. (23). It is then
straightforward to arrive at
\begin{eqnarray}
(M^\dagger_{\rm D} M^{}_{\rm D})^{}_{11} & \simeq &
M^2_1 (R^\dagger R)^{}_{11} \simeq M^2_1
\left(s^2_{14} + s^2_{24} + s^2_{34} \right) \; , \nonumber \\
(M^\dagger_{\rm D} M^{}_{\rm D})^{}_{22} & \simeq &
M^2_2 (R^\dagger R)^{}_{22} \simeq M^2_2
\left(s^2_{15} + s^2_{25} + s^2_{35} \right) \; , \nonumber \\
(M^\dagger_{\rm D} M^{}_{\rm D})^{}_{33} & \simeq &
M^2_3 (R^\dagger R)^{}_{33} \simeq M^2_3
\left(s^2_{16} + s^2_{26} + s^2_{36} \right) \; ;
\end{eqnarray}
and
\begin{eqnarray}
(M^\dagger_{\rm D} M^{}_{\rm D})^{}_{12} & \simeq &
M^{}_1 M^{}_2 (R^\dagger R)^{}_{12} \simeq M^{}_1 M^{}_2
\left(\hat{s}^{}_{14} \hat{s}^*_{15} + \hat{s}^{}_{24} \hat{s}^*_{25}
+ \hat{s}^{}_{34} \hat{s}^*_{35} \right) \; , \nonumber \\
(M^\dagger_{\rm D} M^{}_{\rm D})^{}_{13} & \simeq &
M^{}_1 M^{}_3 (R^\dagger R)^{}_{13} \simeq M^{}_1 M^{}_3
\left(\hat{s}^{}_{14} \hat{s}^*_{16} + \hat{s}^{}_{24} \hat{s}^*_{26}
+ \hat{s}^{}_{34} \hat{s}^*_{36} \right) \; , \nonumber \\
(M^\dagger_{\rm D} M^{}_{\rm D})^{}_{23} & \simeq &
M^{}_2 M^{}_3 (R^\dagger R)^{}_{23} \simeq M^{}_2 M^{}_3
\left(\hat{s}^{}_{15} \hat{s}^*_{16} + \hat{s}^{}_{25} \hat{s}^*_{26}
+ \hat{s}^{}_{35} \hat{s}^*_{36} \right) \; .
\end{eqnarray}
So the unflavored CP-violating asymmetry $\varepsilon^{}_i$ depends
on nine phase differences $\delta^{}_{i4} - \delta^{}_{i5}$,
$\delta^{}_{i4} - \delta^{}_{i6}$ and $\delta^{}_{i5} - \delta^{}_{i6}$
(for $i=1,2,3$), but only six of them are independent.
Because $V \simeq V^{}_0$ holds in the same
approximations as made above, we conclude that there is {\it not} any
direct relationship between the CP violation at low energies (governed
by $\delta^{}_{12}$, $\delta^{}_{13}$ and $\delta^{}_{23}$) and the
unflavored leptogenesis at high energies. As for the flavored leptogenesis,
the relevant CP-violating asymmetries $\varepsilon^{}_{i\alpha}$
also rely on the aforementioned nine phase differences. This
point can be clearly seen from
\begin{eqnarray}
{\rm Im} \left[ (M^*_{\rm D})^{}_{\alpha i} (M^{}_{\rm D})^{}_{\alpha j}
(M^\dagger_{\rm D} M^{}_{\rm D})^{}_{ij} \right] & \simeq &
M^2_i M^2_j ~{\rm Im} \left[ R^*_{\alpha i}
R^{}_{\alpha j} (R^\dagger R)^{}_{ij} \right] \; , \nonumber \\
{\rm Im} \left[ (M^*_{\rm D})^{}_{\alpha i} (M^{}_{\rm D})^{}_{\alpha j}
(M^\dagger_{\rm D} M^{}_{\rm D})^{*}_{ij} \right] & \simeq &
M^2_i M^2_j ~{\rm Im} \left[ R^*_{\alpha i}
R^{}_{\alpha j} (R^\dagger R)^{*}_{ij} \right] \; .
\end{eqnarray}
Given the masses of three heavy Majorana neutrinos and the flavor mixing
parameters between light and heavy Majorana neutrinos, it is then
possible to determine $\varepsilon^{}_{i\alpha}$ and $\varepsilon^{}_i$
(for $i=1,2,3$ and $\alpha = e, \mu, \tau$) so as to realize the flavored
or unflavored leptogenesis mechanism.

\section{Summary and remarks}

An appealing and puzzling feature of the standard model is that it
happens to have three species of leptons and quarks. If extra species of
matter particles exist, no matter whether they are sequential in or
exotic to the standard model itself, they definitely signify new physics.
In this paper
we have conjectured the presence of three species of sterile neutrinos
and presented a full parametrization of the $6\times 6$ flavor mixing
matrix for active and sterile neutrinos in terms of fifteen rotation angles
and fifteen phase angles. Such an exercise makes sense because we {\it do}
have some preliminary observational hints on light sterile neutrinos,
and the theoretical motivation for the existence of heavy Majorana neutrinos
in the seesaw and leptogenesis mechanisms is also very strong.

We have shown that this standard parametrization of the $6\times 6$
flavor mixing matrix in the (3+3) scenario of active and sterile neutrino
mixing allows us to clearly describe the salient features of some problems
in neutrino phenomenology. Four examples have been briefly discussed
in this connection: (a) possible
contributions of light sterile neutrinos to the tritium beta decay
and neutrinoless double-beta decay; (b) leptonic CP violation
and deformed unitarity triangles of the $3\times 3$ flavor
mixing matrix $V$ of three active neutrinos; (c) a reconstruction of the
$6\times 6$ neutrino mass matrix in the type-(I+II) seesaw mechanism;
and (d) flavored and unflavored leptogenesis scenarios in the type-I
seesaw mechanism with three heavy Majorana neutrinos. Our results are
expected to be useful to understand the impacts of extra neutrino species
on the standard weak interactions and neutrino oscillations in a better
and more straightforward way. Furthermore, the parametrization itself
can also be applied to the quark sector if extra species of quarks are
conjectured.

Let us stress that the presence of new degrees of freedom in the neutrino
sector may apparently violate the unitarity of the $3\times 3$ flavor
mixing matrix $V$ of three active neutrinos in the weak charged-current
interactions. Hence testing the unitarity of $V$ in neutrino oscillations
and searching for the signatures of new neutrinos at TeV-scale
colliders can be complementary to each other, both qualitatively and
quantitatively, in order to deeply understand the intrinsic
properties of Majorana particles. On the other hand, light or heavy
sterile neutrinos can have significant consequences in cosmology. The
latter provides us with a good playground to study hot dark matter in
the presence of sub-eV sterile neutrinos and warm dark matter in the form
of keV sterile neutrinos, together with the cosmological matter-antimatter
asymmetry via leptogenesis in which heavy Majorana neutrinos and their
decays play the key role. We hope that any experimental breakthrough in
these aspects will pave the way towards the true theory
of massive neutrinos.

\vspace{0.5cm}

The author would like to thank B. Dziewit, Y.F. Li, W. Rodejohann,
S. Su, S. Zhou and M. Zralek for useful discussions.
This work was supported in part by the National Natural Science Foundation
of China under grant No. 11135009.

\newpage
\appendix
\section{}

The $6\times 6$ unitary matrix $\cal U$ in Eq. (1) can be expressed
as a product of fifteen two-dimensional rotation matrices in a
six-dimensional complex space \cite{Xing08}:
\begin{eqnarray}
{\cal U} = \left(O^{}_{56} O^{}_{46} O^{}_{36} O^{}_{26}
O^{}_{16}\right) \left(O^{}_{45} O^{}_{35} O^{}_{25}
O^{}_{15}\right) \left(O^{}_{34} O^{}_{24} O^{}_{14}\right)
\left(O^{}_{23} O^{}_{13}\right) O^{}_{12} \; ,
\end{eqnarray}
where $O^{}_{ij}$ (for $1\leq i < j \leq 6$) are unitary and read as
follows:
\begin{eqnarray}
O^{}_{12} & = & \left( \matrix{c^{}_{12} & \hat{s}^*_{12} & 0 & 0 &
0 & 0 \cr -\hat{s}^{}_{12} & c^{}_{12} & 0 & 0 & 0 & 0 \cr 0 & 0 & 1
& 0 & 0 & 0 \cr 0 & 0 & 0 & 1 & 0 & 0 \cr 0 & 0 & 0 & 0 & 1 & 0 \cr
0 & 0 & 0 & 0 & 0 & 1 \cr} \right) \; ;
\end{eqnarray}
and
\begin{eqnarray}
O^{}_{13} & = & \left( \matrix{c^{}_{13} & 0 & \hat{s}^*_{13}
& 0 & 0 & 0 \cr 0 & 1 & 0 & 0 & 0 & 0 \cr -\hat{s}^{}_{13} & 0 &
c^{}_{13} & 0 & 0 & 0 \cr 0 & 0 & 0 & 1 & 0 & 0 \cr 0 & 0 & 0 & 0
& 1 & 0 \cr 0 & 0 & 0 & 0 & 0 & 1 \cr} \right) \; ,
\nonumber \\
O^{}_{23} & = & \left( \matrix{1 & 0 & 0 & 0 & 0 & 0 \cr 0 &
c^{}_{23} & \hat{s}^*_{23} & 0 & 0 & 0 \cr 0 & -\hat{s}^{}_{23} &
c^{}_{23} & 0 & 0 & 0 \cr 0 & 0 & 0 & 1 & 0 & 0 \cr 0 & 0 & 0 & 0 &
1 & 0 \cr 0 & 0 & 0 & 0 & 0 & 1 \cr} \right) \; ;
\end{eqnarray}
and
\begin{eqnarray}
O^{}_{14} & = & \left( \matrix{c^{}_{14} & 0 & 0 & \hat{s}^*_{14} &
0 & 0 \cr 0 & 1 & 0 & 0 & 0 & 0 \cr 0 & 0 & 1 & 0 & 0 & 0 \cr
-\hat{s}^{}_{14} & 0 & 0 & c^{}_{14} & 0 & 0 \cr 0 & 0 & 0 & 0 & 1 &
0 \cr 0 & 0 & 0 & 0 & 0 & 1 \cr} \right) \; ,
\nonumber \\
O^{}_{24} & = & \left( \matrix{1 & 0 & 0 & 0 & 0 & 0 \cr 0 &
c^{}_{24} & 0 & \hat{s}^*_{24} & 0 & 0 \cr 0 & 0 & 1 & 0 & 0 & 0 \cr
0 & -\hat{s}^{}_{24} & 0 & c^{}_{24} & 0 & 0 \cr 0 & 0 & 0 & 0 & 1 &
0 \cr 0 & 0 & 0 & 0 & 0 & 1 \cr} \right) \; ,
\nonumber \\
O^{}_{34} & = & \left( \matrix{1 & 0 & 0 & 0 & 0 & 0 \cr 0 & 1 & 0 &
0 & 0 & 0 \cr 0 & 0 & c^{}_{34} & \hat{s}^*_{34} & 0 & 0 \cr 0 & 0 &
-\hat{s}^{}_{34} & c^{}_{34} & 0 & 0 \cr 0 & 0 & 0 & 0 & 1 & 0 \cr 0
& 0 & 0 & 0 & 0 & 1 \cr} \right) \; ;
\end{eqnarray}
and
\begin{eqnarray}
O^{}_{15} & = & \left( \matrix{c^{}_{15} & 0 & 0 & 0 &
\hat{s}^*_{15} & 0 \cr 0 & 1 & 0 & 0 & 0 & 0 \cr 0 & 0 & 1 & 0 & 0 &
0 \cr 0 & 0 & 0 & 1 & 0 & 0 \cr -\hat{s}^{}_{15} & 0 & 0 & 0 &
c^{}_{15} & 0 \cr 0 & 0 & 0 & 0 & 0 & 1 \cr} \right) \; ,
\nonumber \\
O^{}_{25} & = & \left( \matrix{1 & 0 & 0 & 0 & 0 & 0 \cr 0 &
c^{}_{25} & 0 & 0 & \hat{s}^*_{25} & 0 \cr 0 & 0 & 1 & 0 & 0 & 0 \cr
0 & 0 & 0 & 1 & 0 & 0 \cr 0 & -\hat{s}^{}_{25} & 0 & 0 & c^{}_{25} &
0 \cr 0 & 0 & 0 & 0 & 0 & 1 \cr} \right) \; ,
\nonumber \\
O^{}_{35} & = & \left( \matrix{1 & 0 & 0 & 0 & 0 & 0 \cr 0 & 1 & 0 &
0 & 0 & 0 \cr 0 & 0 & c^{}_{35} & 0 & \hat{s}^*_{35} & 0 \cr 0 & 0 &
0 & 1 & 0 & 0 \cr 0 & 0 & -\hat{s}^{}_{35} & 0 & c^{}_{35} & 0 \cr 0
& 0 & 0 & 0 & 0 & 1 \cr} \right) \; ,
\nonumber \\
O^{}_{45} & = & \left( \matrix{1 & 0 & 0 & 0 & 0 & 0 \cr 0 & 1 & 0 &
0 & 0 & 0 \cr 0 & 0 & 1 & 0 & 0 & 0 \cr 0 & 0 & 0 & c^{}_{45} &
\hat{s}^*_{45} & 0 \cr 0 & 0 & 0 & -\hat{s}^{}_{45} & c^{}_{45} & 0
\cr 0 & 0 & 0 & 0 & 0 & 1 \cr} \right) \; ;
\end{eqnarray}
and
\begin{eqnarray}
O^{}_{16} & = & \left( \matrix{c^{}_{16} & 0 & 0 & 0 & 0 &
\hat{s}^*_{16} \cr 0 & 1 & 0 & 0 & 0 & 0 \cr 0 & 0 & 1 & 0 & 0 & 0
\cr 0 & 0 & 0 & 1 & 0 & 0 \cr 0 & 0 & 0 & 0 & 1 & 0 \cr
-\hat{s}^{}_{16} & 0 & 0 & 0 & 0 & c^{}_{16} \cr} \right) \; ,
\nonumber \\
O^{}_{26} & = & \left( \matrix{1 & 0 & 0 & 0 & 0 & 0 \cr 0 &
c^{}_{26} & 0 & 0 & 0 & \hat{s}^*_{26} \cr 0 & 0 & 1 & 0 & 0 & 0 \cr
0 & 0 & 0 & 1 & 0 & 0 \cr 0 & 0 & 0 & 0 & 1 & 0 \cr 0 &
-\hat{s}^{}_{26} & 0 & 0 & 0 & c^{}_{26} \cr} \right) \; ,
\nonumber \\
O^{}_{36} & = & \left( \matrix{1 & 0 & 0 & 0 & 0 & 0 \cr 0 & 1 & 0 &
0 & 0 & 0 \cr 0 & 0 & c^{}_{36} & 0 & 0 & \hat{s}^*_{36} \cr 0 & 0 &
0 & 1 & 0 & 0 \cr 0 & 0 & 0 & 0 & 1 & 0 \cr 0 & 0 & -\hat{s}^{}_{36}
& 0 & 0 & c^{}_{36} \cr} \right) \; ,
\nonumber \\
O^{}_{46} & = & \left( \matrix{1 & 0 & 0 & 0 & 0 & 0 \cr 0 & 1 & 0 &
0 & 0 & 0 \cr 0 & 0 & 1 & 0 & 0 & 0 \cr 0 & 0 & 0 & c^{}_{46} & 0 &
\hat{s}^*_{46} \cr 0 & 0 & 0 & 0 & 1 & 0 \cr 0 & 0 & 0 &
-\hat{s}^{}_{46} & 0 & c^{}_{46} \cr} \right) \; ,
\nonumber \\
O^{}_{56} & = & \left( \matrix{1 & 0 & 0 & 0 & 0 & 0 \cr 0 & 1 & 0 &
0 & 0 & 0 \cr 0 & 0 & 1 & 0 & 0 & 0 \cr 0 & 0 & 0 & 1 & 0 & 0 \cr 0
& 0 & 0 & 0 & c^{}_{56} & \hat{s}^*_{56} \cr 0 & 0 & 0 & 0 &
-\hat{s}^{}_{56} & c^{}_{56} \cr} \right) \; .
\end{eqnarray}
In the above equations we have defined $c^{}_{ij} \equiv
\cos\theta^{}_{ij}$ and $\hat{s}^{}_{ij} \equiv e^{i\delta^{}_{ij}}
\sin\theta^{}_{ij}$ with $\theta^{}_{ij}$ and $\delta^{}_{ij}$ being
the rotation angle and phase angle, respectively. Because some of
the two-dimensional rotation matrices can commute with each other,
it is possible to rewrite Eq. (A1) as
\begin{eqnarray}
{\cal U} = \left(O^{}_{56} O^{}_{46} O^{}_{45}\right)
\left(O^{}_{36} O^{}_{26} O^{}_{16} O^{}_{35} O^{}_{25} O^{}_{15}
O^{}_{34} O^{}_{24} O^{}_{14}\right) \left(O^{}_{23} O^{}_{13}
O^{}_{12}\right) \; .
\end{eqnarray}
On the right-hand side of Eq. (A7), the combination $(O^{}_{23}
O^{}_{13} O^{}_{12})$ purely describes the flavor mixing among three
active neutrinos while $(O^{}_{56} O^{}_{46} O^{}_{45})$ purely
describes the flavor mixing among three sterile neutrinos. It is the
combination $(O^{}_{36} \cdots O^{}_{14})$ in Eq. (A7) that allows
the active and sterile sectors to ``talk" to each other, as
discussed in section II.

\section{}

Given the $n\times n$ flavor mixing matrix, one may always calculate
its $(n-1)^2 (n-2)^2/4$ rephasing invariants of CP violation, the
so-called Jarlskog parameters \cite{J}. As for the $6\times 6$
unitary matrix $\cal U$ under discussion, we totally have 100
invariants of this nature. But we are mainly concerned about the
Jarlskog parameters of $V \equiv A V^{}_0$ defined in section II:
\begin{eqnarray}
J^{ij}_{\alpha\beta} \equiv {\rm Im}(V^{}_{\alpha i} V^{}_{\beta j}
V^*_{\alpha j} V^*_{\beta i}) \; ,
\end{eqnarray}
in which the Greek indices run over $(e, \mu, \tau)$ and the Latin
indices run over $(1,2,3$). Since $V$ describes the flavor mixing of
three active neutrinos, $J^{ij}_{\alpha\beta}$ measure the
corresponding CP-violating effects in their oscillations. In the
absence of three sterile neutrinos (i.e., $A = {\bf 1}$ and $V =
V^{}_0$), one arrives at a universal Jarlskog parameter
\begin{eqnarray}
J^{}_0 \equiv J^{12}_{e \mu} = J^{23}_{e \mu} = J^{31}_{e \mu} =
J^{12}_{\mu \tau} = J^{23}_{\mu \tau} = J^{31}_{\mu \tau} =
J^{12}_{\tau e} = J^{23}_{\tau e} = J^{31}_{\tau e} = c^{}_{12}
s^{}_{12} c^2_{13} s^{}_{13} c^{}_{23} s^{}_{23} \sin\delta
\end{eqnarray}
with $\delta \equiv \delta^{}_{13} - \delta^{}_{12} -
\delta^{}_{23}$, as guaranteed by the unitarity of $V^{}_0$. When
the contributions of three sterile neutrinos are switched on,
$J^{ij}_{\alpha\beta}$ can be calculated with the help of Eqs. (5)
and (12) in a good approximation. One finds
\begin{eqnarray}
J^{12}_{e\mu} & \simeq & J^{}_0 + c^{}_{12} s^{}_{12} c^{}_{23} {\rm
Im} X \; , \nonumber \\
J^{12}_{\tau e} & \simeq & J^{}_0 + c^{}_{12} s^{}_{12} s^{}_{23}
{\rm Im} Y \; , \nonumber \\
J^{12}_{\mu \tau} & \simeq & J^{}_0 + c^{}_{12} s^{}_{12} c^{}_{23}
s^{}_{23} \left( s^{}_{23} {\rm Im} X + c^{}_{23} {\rm Im} Y \right)
\; , \nonumber \\
J^{23}_{\mu \tau} & \simeq & J^{}_0 + c^{}_{12} c^{}_{23} s^{}_{23}
\left( s^{}_{12} s^{}_{23} {\rm Im} X + s^{}_{12} c^{}_{23} {\rm Im}
Y + c^{}_{12} {\rm Im} Z \right) \; , \nonumber \\
J^{31}_{\mu \tau} & \simeq & J^{}_0 + s^{}_{12} c^{}_{23} s^{}_{23}
\left( c^{}_{12} s^{}_{23} {\rm Im} X + c^{}_{12} c^{}_{23} {\rm Im}
Y - s^{}_{12} {\rm Im} Z \right) \; ,
\end{eqnarray}
and $J^{23}_{e\mu} \simeq J^{31}_{e\mu} \simeq J^{23}_{\tau e}
\simeq J^{31}_{\tau e} \simeq J^{}_0$ \cite{Wu}, where $X \equiv
{\cal X} e^{-i\delta^{}_{12}}$, $Y \equiv {\cal Y} e^{-i
(\delta^{}_{12} + \delta^{}_{23})}$ and $Z \equiv {\cal Z}
e^{-i\delta^{}_{23}}$ with $\cal X$, $\cal Y$ and $\cal Z$ being
defined below Eq. (17). Note that we have assumed $\theta^{}_{13}$,
$\theta^{}_{i4}$, $\theta^{}_{i5}$ and $\theta^{}_{i6}$ (for
$i=1,2,3$) to be small in our calculations, and thus the terms of
${\cal O}(s^{}_{13} |X|)$, ${\cal O}(s^{}_{13} |Y|)$ and ${\cal
O}(s^{}_{13} |Z|)$ together with those higher-order terms have been
omitted from the above results. The fact that $J^{23}_{e\mu} \simeq
J^{31}_{e\mu} \simeq J^{23}_{\tau e} \simeq J^{31}_{\tau e} \simeq
J^{}_0$ holds in the above approximation is simply because they all
involve the smallest matrix element of $V$ (i.e., $V^{}_{e3} \simeq
\hat{s}^*_{13}$) \cite{Malinsky}.

It is well known that the maximal value of $J^{}_0$ is $J^{\rm
max}_0 = 1/(6\sqrt{3}) \simeq 9.6\%$ \cite{J}. In comparison, the
magnitudes of ${\rm Im} X$, ${\rm Im} Y$ and ${\rm Im} Z$ are likely
to reach the percent level if those active-sterile mixing angles are
of ${\cal O}(0.1)$ and the relevant CP-violating phases are of
${\cal O}(1)$. So the five Jarlskog parameters in Eq. (B3) might
deviate from the standard one $J^{}_0$ in a significant way,
depending on the constructive or destructive contributions from
three sterile neutrinos. It is therefore important to observe all
the nine Jarlskog parameters in neutrino oscillations. Taking
account of the non-unitarity of $V$, one may easily derive the
probabilities of $\nu^{}_\alpha \to \nu^{}_\beta$ and
$\overline{\nu}^{}_\alpha \to \overline{\nu}^{}_\beta$ oscillations
\cite{Xing08,Antusch}:
\begin{eqnarray}
P(\nu^{}_\alpha \rightarrow \nu^{}_\beta) & = & \frac{\displaystyle
\sum^{}_i |V^{}_{\alpha i}|^2 |V^{}_{\beta i}|^2 + 2 \sum^{}_{i<j}
{\rm Re} \left( V^{}_{\alpha i} V^{}_{\beta j} V^*_{\alpha j}
V^*_{\beta i} \right) \cos \Delta^{}_{ij} - 2 \sum^{}_{i<j}
J^{ij}_{\alpha\beta} \sin\Delta^{}_{ij}}{\displaystyle \left(
VV^\dagger\right)^{}_{\alpha\alpha} \left(
VV^\dagger\right)^{}_{\beta\beta}} \; , \nonumber \\
P(\overline{\nu}^{}_\alpha \rightarrow \overline{\nu}^{}_\beta) & =
& \frac{\displaystyle \sum^{}_i |V^{}_{\alpha i}|^2 |V^{}_{\beta
i}|^2 + 2 \sum^{}_{i<j} {\rm Re} \left( V^{}_{\alpha i} V^{}_{\beta
j} V^*_{\alpha j} V^*_{\beta i} \right) \cos \Delta^{}_{ij} + 2
\sum^{}_{i<j} J^{ij}_{\alpha\beta} \sin\Delta^{}_{ij}}{\displaystyle
\left( VV^\dagger\right)^{}_{\alpha\alpha} \left(
VV^\dagger\right)^{}_{\beta\beta}} \; ,
\end{eqnarray}
where $\Delta^{}_{ij} \equiv \Delta m^2_{ij} L/(2E)$ with $\Delta
m^2_{ij} \equiv m^2_i - m^2_j$, $E$ being the neutrino beam energy
and $L$ being the baseline length. As a consequence,
\begin{eqnarray}
{\cal A}^{}_{\alpha\beta} \equiv P(\nu^{}_\alpha \rightarrow
\nu^{}_\beta) - P(\overline{\nu}^{}_\alpha \rightarrow
\overline{\nu}^{}_\beta) & = & \frac{4}{\displaystyle \left(
AA^\dagger\right)^{}_{\alpha\alpha} \left(
AA^\dagger\right)^{}_{\beta\beta}} \sum^{}_{i<j}
J^{ij}_{\alpha\beta} \sin\Delta^{}_{ij} \nonumber \\
& \simeq & 4 \left[ J^{12}_{\alpha\beta} \sin\Delta^{}_{21} + \left(
J^{13}_{\alpha\beta} + J^{23}_{\alpha\beta} \right)
\sin\Delta^{}_{32} \right] \; ,
\end{eqnarray}
where $AA^\dagger \simeq {\bf 1}$ and $\Delta^{}_{31} \simeq
\Delta^{}_{32}$ (i.e., $\Delta m^2_{31} \simeq \Delta m^2_{32}$
\cite{PDG}) have been taken into account. This result implies that
both $J^{12}_{\alpha\beta}$ and $J^{13}_{\alpha\beta} +
J^{23}_{\alpha\beta}$ can in principle be determined if the
baseline of neutrino oscillations is sufficiently long. Of course,
terrestrial matter effects may more or less contaminate the genuine
CP-violating effects in such long-baseline neutrino oscillation
experiments and should be properly treated \cite{Non}.

\newpage

\begin{figure}[t]
\vspace{0cm}
\epsfig{file=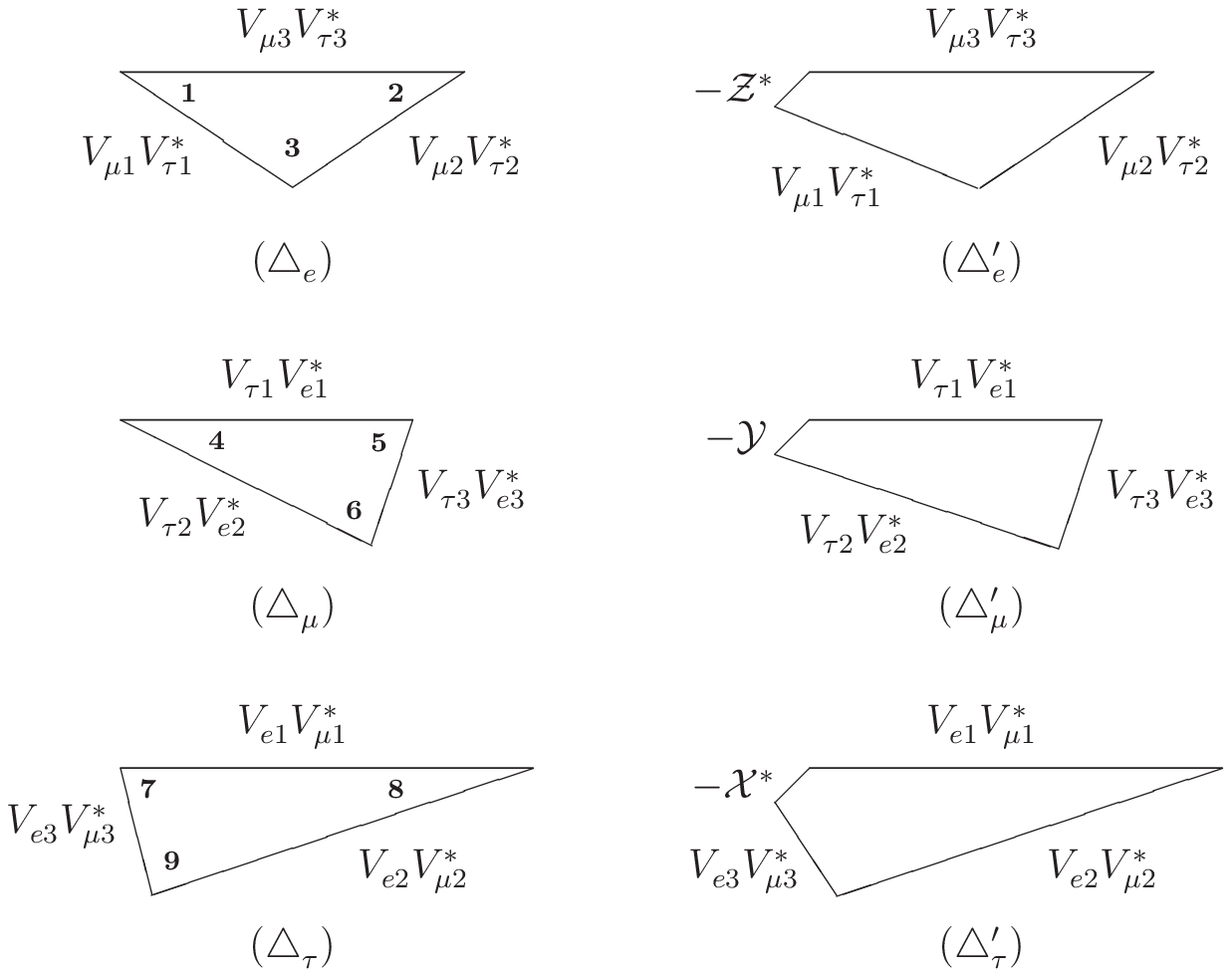,bbllx=4.6cm,bblly=14cm,bburx=13.4cm,bbury=29cm,%
width=10cm,height=16cm,angle=0,clip=0} \vspace{-0.5cm} \caption{{\bf
Left panel:} three unitarity triangles of $V \equiv A V^{}_0$ in the
absence of three sterile neutrinos (i.e., $A = {\bf 1}$ and $V =
V^{}_0$); {\bf Right panel:} three deformed unitarity triangles of
$V \equiv A V^{}_0$ in the presence of three sterile neutrinos,
where $\cal X$, $\cal Y$ and $\cal Z$ are defined below Eq. (17).}
\end{figure}


\begin{thebibliography}{99}

\bibitem{SS} P. Minkowski, Phys. Lett. B {\bf 67}, 421 (1977);
T. Yanagida, in {\it Proceedings of the Workshop on Unified Theory
and the Baryon Number of the Universe}, edited by O. Sawada and A.
Sugamoto (KEK, Tsukuba, 1979); M. Gell-Mann, P. Ramond, and R.
Slansky, in {\it Supergravity}, edited by P. van Nieuwenhuizen and
D. Freedman (North Holland, Amsterdam, 1979); S.L. Glashow, in
{\it Quarks and Leptons}, edited by M. L$\acute{\rm e}$vy {\it et
al.} (Plenum, New York, 1980); R.N. Mohapatra and G. Senjanovic,
Phys. Rev. Lett. {\bf 44}, 912 (1980).

\bibitem{FY} M. Fukugita and T. Yanagida, Phys. Lett. B {\bf 174},
45 (1986).

\bibitem{LSND} A. Aguilar {\it et al.} (LSND Collaboration),
Phys. Rev. D {\bf 64}, 112007 (2001).

\bibitem{M} A.A. Aguilar-Arevalo {\it et al.} (MiniBooNE Collaboration),
Phys. Rev. Lett. {\bf 105}, 181801 (2010).

\bibitem{R} G. Mention {\it et al.}, Phys. Rev. D {\bf  83}, 073006
(2011).

\bibitem{Schwetz} J. Kopp, M. Maltoni, and T. Schwetz,
Phys. Rev. Lett. {\bf 107}, 091801 (2011);
C. Giunti and M. Laveder, Phys. Rev. D {\bf 84}, 073008 (2011).

\bibitem{Raffelt} J. Hamann {\it et al.}, Phys. Rev. Lett.
{\bf 105}, 181301 (2010); JCAP {\bf 1109}, 034 (2011);
E. Giusarma {\it et al.}, Phys. Rev. D {\bf 83}, 115023 (2011).

\bibitem{Mangano} See, e.g., G. Mangano and P.D. Serpico,
Phys. Lett. B {\bf 701}, 296 (2011); and references therein.

\bibitem{Bode} P. Bode, J.P. Ostriker, and N. Turok,
Astrophys. J. {\bf 556}, 93 (2001).

\bibitem{keV} See, e.g., T. Asaka, S. Blanchet, and M. Shaposhnikov,
Phys. Lett. B {\bf 631}, 151 (2005); A. Kusenko, Phys. Rev. Lett.
{\bf 97}, 241301 (2006); K. Petraki and A. Kusenko, Phys. Rev. D
{\bf 77}, 065014 (2008); A. Kusenko, F. Takahashi, and T. Yanagida,
Phys. Lett. B {\bf 693}, 144 (2010); M. Lindner, A. Merle, and V.
Niro, JCAP {\bf 1101}, 034 (2011).

\bibitem{eV} See, e.g., D.K. Ghosh, G. Senjanovic, and Y. Zhang,
Phys. Lett. B {\bf 698}, 420 (2011); V. Barger, P.F. Perez, and S.
Spinner, Phys. Lett. B {\bf 696}, 509 (2011); J. Barry, W.
Rodejohann, and H. Zhang, JHEP {\bf 1107}, 091 (2011); and
references therein.

\bibitem{Desert} Z.Z. Xing, invited plenary talk given at
COSMO/CosPA 2010, September 2010, Tokyo;
Y.F. Li and Z.Z. Xing, Phys. Lett. B {\bf 695}, 205 (2011).

\bibitem{BF} A formally exact but very different parametrization of
the $6\times 6$ neutrino mixing matrix in the type-I seesaw framework
has been done in: M. Blennow and E. Fernandez-Martinez, Phys. Lett. B
{\bf 704}, 223 (2011).

\bibitem{PDG} Particle Data Group, K. Nakamura {\it et al.},
J. Phys. G {\bf 37}, 075021 (2010).

\bibitem{Xing08} Z.Z. Xing, Phys. Lett. B {\bf 660}, 515 (2008).

\bibitem{MNS} Z. Maki, M. Nakagawa, and S. Sakata, Prog. Theor.
Phys. {\bf 28}, 870 (1962); B. Pontecorvo, Sov. Phys. JETP {\bf 26},
984 (1968).

\bibitem{Antusch} S. Antusch, C. Biggio, E. Fernandez-Martinez, M.B.
Gavela, and J. Lopez-Pavon, JHEP {\bf 0610}, 084 (2006).

\bibitem{MS} See, e.g., M. Maltoni and T. Schwetz, Phys. Rev. D {\bf 76},
093005 (2007).

\bibitem{Xing03} Z.Z. Xing, Phys. Rev. D {\bf 68}, 053002 (2003);
Y. BenTov and A. Zee, Phys. Rev. D {\bf 84}, 073012 (2011).

\bibitem{LL} Y.F. Li and S.S. Liu, arXiv:1110.5795.

\bibitem{Rodejohann} For a recent review,
see: W. Rodejohann, Int. J. Mod. Phys. E {\bf 20}, 1833 (2011).
See, also, W. Rodejohann and J.W.F. Valle,
Phys. Rev. D {\bf 84}, 073011 (2011).

\bibitem{FX99} H. Fritzsch and Z.Z. Xing, Prog. Part. Nucl. Phys. {\bf 45}, 1
(2000).

\bibitem{J} C. Jarlskog, Phys. Rev. Lett. {\bf 55}, 1039 (1985).

\bibitem{Wu} X.G. Wu, thesis for a bachelor's degree, Peking
University (2009).

\bibitem{Malinsky} M. Malinsky, T. Ohlsson, Z.Z. Xing, and H. Zhang,
Phys. Lett. B {\bf 679}, 242 (2009).

\bibitem{Non} E. Fernandez-Martinez, M.B. Gavela, J.
L$\rm\acute{o}$pez-Pav$\rm\acute{o}$n, and O. Yasuda, Phys. Lett. B
{\bf 649}, 427 (2007); Z.Z. Xing, Phys. Lett. B {\bf 660}, 515
(2008); S. Luo, Phys. Rev. D {\bf 78}, 016006 (2008); S. Goswami and
T. Ota, Phys. Rev. D {\bf 78}, 033012 (2008); G. Altarelli and D.
Meloni, Nucl. Phys. B {\bf 809}, 158 (2009); Z.Z. Xing, Prog. Theor.
Phys. Suppl. {\bf 180}, 112 (2009); S. Antusch, M. Blennow, E.
Fernandez-Martinez, and J. L$\rm\acute{o}$pez-Pav$\rm\acute{o}$n,
Phys. Rev. D {\bf 80}, 033002 (2009); M. Malinsky, T. Ohlsson, Z.Z.
Xing, and H. Zhang, Phys. Lett. B {\bf 679}, 242 (2009).

\bibitem{T2}
J. Schechter and J.W.F. Valle, Phys. Rev. D {\bf 22}, 2227 (1980);
{\bf 25}, 774 (1982);
T.P. Cheng and L.F. Li, Phys. Rev. D {\bf 22}, 2860 (1980);
M. Magg and C. Wetterich, Phys. Lett. B {\bf 94}, 61 (1980);
G. Lazarides, Q. Shafi, and C. Wetterich, Nucl. Phys. B {\bf 181}, 287 (1981);
R.N. Mohapatra and G. Senjanovic, Phys. Rev. D {\bf 23}, 165 (1981).

\bibitem{Zralek} A numerical reconstruction of the $3\times 3$
Majorana neutrino mass matrix has been done by B. Dziewit, K. Kajda,
J. Gluza, and M. Zralek, Phys. Rev. D {\bf 74}, 033003 (2006). In
principle, it is also possible to numerically reconstruct the
$6\times 6$ neutrino mass matrix in the type-I or type-(I+II)
seesaw mechanism (or equivalently its $3\times 3$
submatrices $M^{}_{\rm L}$, $M^{}_{\rm D}$ and $M^{}_{\rm R}$).

\bibitem{Xing10} Z.Z. Xing, Chin. Phys. C {\bf 34}, 1 (2010).

\bibitem{Book} For a detailed calculation of $\varepsilon^{}_{i\alpha}$,
see: Z.Z. Xing and S. Zhou, {\it Neutrinos in Particle
Physics, Astronomy and Cosmology} (Zhejiang University Press and
Springer-Verlag, 2011). The original derivation of $\varepsilon^{}_{i\alpha}$
can be found in: T. Endoh, T. Morozumi, and Z.H. Xiong,
Prog. Theor. Phys. {\bf 111}, 123 (2004).

\end{thebibliography}
\end{document}